# Performance of Underwater Quantum Key Distribution with Polarization Encoding

SHI-CHENG ZHAO, XIN-HONG HAN, YA XIAO, YUAN SHEN, YONG-JIAN GU* AND WEN-DONG LI*

*Department of Physics, Ocean University of China, Qingdao 266100, China.*

E-mail: : yjgu@ouc.edu.cn, liwd@ouc.edu.cn

**Abstract:** Underwater quantum key distribution (QKD) has potential applications in absolutely secure underwater communication. However, the performance of underwater QKD is limited by the optical elements, background light, and dark counts of the detector. In this paper, we propose a modified formula for the quantum bit error rate (QBER), which takes into account the effect of detector efficiency on the QBER caused by the background light. Then we calculate the QBER of the polarization encoding BB84 protocol in Jerlov type seawater by analyzing the effect of the background light and optical components in a more realistic situation. Finally, we further analyze the final key rate and the maximum secure communication distance in three propagation modes, i.e., upward, downward and horizontal modes. We find that secure QKD can be performed in the clearest Jerlov type seawater at a distance of hundreds of meters, even in the worst downward propagation mode. Specifically, by optimizing the system parameters, it is possible to securely transmit information with a rate of 67kbits/s at a distance of 100 m in the seawater channel with an attenuation coefficient of 0.03/m at night. For practical underwater QKD, the performance can also be improved by using decoy states. Our results are useful to long distance underwater quantum communication.

## 1. Introduction

Underwater communication is vital for underwater sensor networks, submarines, and all types of underwater vehicles and it can be made secure using quantum key distribution (QKD). QKD enables two remote parties to set up secure keys whose security is based on the basic physical properties of quantum states, rather than relying on the computational intractability of certain mathematical functions in traditional cryptography. In1984, Bennett and Brassard proposed the first QKD protocol [1], the absolute security of which has been proved using one-time pad encryption [2–4]. Since the first QKD experiment in 1989 [5] with a distance of 32 cm, a strong research effort has been devoted to achieving practical QKD. Great progress has been made on QKD in free space and optical fiber [6–8]. In 2016, a low-earth-orbit satellite to implement decoy state QKD was launched [9] and successfully realized satellite-to-ground QKD over a distance of 1200km with a key rate above the kilohertz level [10, 11].

However, little progress has been made on underwater QKD in spite of the following works. |RUnderwater QKD was first proposed in 2012; Ref. [12, 13] showed that underwater QKD can be performed at a distance of 100m, thus proved the feasibility of underwater quantum communication . In 2014, Monte Carlo simulation was used to study the propagation characteristics of polarized photons in seawater [14], and the effect of the underwater channel on QKD was analyzed. The results in Ref. [14] showed that underwater QKD can be performed with a sifted bit rate of 45kb/s at the communication distance of 107 m in Jerlov type-I seawater in the environmental condition of starlight only. These studies showed that secure QKD can be achieved with a distance of one hundred meters in the clearest

seawater in theory. Ji et al. completed the first experiment in underwater quantum communications through a Jerlov type seawater channel [15], which showed polarization qubit and entanglement can maintain well after going through seawater channel and therefore experimentally confirmed the feasibility of underwater quantum communication. The influence of imperfect optical elements on the polarization states for polarization encoding underwater QKD has never been analyzed, and the influence of background light on the quantum bit error rate (QBER) is not clear.

In this paper, the QBER and final key rate are calculated by analyzing the effect of the background light and optical components using a modified QBER formula in three propagation modes (upward, downward, and horizontal). We first modify the formula for calculating the QBER for a QKD system, because the previous formula of the QBER ignores the influence of detection efficiency and optical element transmittance on the background light. Then we calculate the background light of the underwater channel using Hydrolight [16], a professional software that uses Fortran and invariant imbedding, to calculate the background light in oceanic optics, and analyze the effect of the optical elements in a typical underwater BB84 system with polarization coding. We investigate the QBER to evaluate the performance of underwater QKD and the main factors that affect the QBER of underwater QKD according to the modified formula. Finally, we calculate the sifted key rate and final key rate for underwater QKD in the clearest Jerlov type seawater in three propagation modes (upward, downward, and horizontal) under full moon condition. The results show that the final key rate can reach tens of kbits/s for the downward and horizontal modes at a distance of 100m, and the maximum secure distance can reach hundreds of meters even in the worst downward propagation mode. Specifically, by optimizing the system parameters, the secure key rate for QKD can reach 67kbits/s at a distance of 100 m in the seawater channel with an attenuation coefficient 0.03/m. For practical underwater QKD, the performance can also be improved if by using decoy states[17, 18].

## 2. QBER OF BB84 PROTOCOL

According to Ref. [19-21], when Alice transmits the weak coherent pulses to Bob, the QBER reads

$$QBER = \frac{R_{wrong}}{R_{total}} = Q_{opt} + Q_{I_{dc}} + Q_{bac}$$

$$Q_{opt} = \frac{P\frac{\mu\eta}{2\Delta t}e^{-\chi_c r}}{\frac{\mu\eta}{\%2\Delta t}e^{-\chi_c r} + 2I_{dc} + \frac{LA\Delta t^{'}\lambda\Delta\lambda\Omega}{2hc\Delta t}}$$

$$Q_{I_{dc}} = \frac{I_{dc}}{\frac{\mu\eta}{\%2\Delta t}e^{-\chi_c r} + 2I_{dc} + \frac{LA\Delta t^{'}\lambda\Delta\lambda\Omega}{2hc\Delta t}}$$  (1)

$$Q_{bac} = \frac{\frac{LA\Delta t^{'}\lambda\Delta\lambda\Omega}{4hc\Delta t}}{\frac{\mu\eta}{\%2\Delta t}e^{-\chi_c r} + 2I_{dc} + \frac{LA\Delta t^{'}\lambda\Delta\lambda\Omega}{2hc\Delta t}}$$

where $R_{wrong}$ is the ratio of wrong bits, including the wrong bits induced by the imperfection of optical elements ($P\frac{\mu\eta}{2\Delta t}e^{-\chi_c r}$), dark counts and background light. $R_{total}$ is the ratio of total bits detected by Bob, including signal $\frac{\mu\eta}{2\Delta t}e^{-\chi_c r}$ dark counts and background light. $Q_{opt}, Q_{I_{dc}}$ and $Q_{bac}$ are the QBERs induced by the imperfection of optical elements, the dark count of the single-photon detectors and the background light, respectively. P is the polarization contrast, which means the ratio of two optical power levels with orthogonal polarization. $\eta$ is the detector efficiency for the signal and background light with the same wavelength as the signal. As the filter is with a narrow bandwidth, $\eta$ is constant within the bandwidth of the filter. $\Delta t$ is the bit period, $\chi_c$ is the attenuation coefficient, r is the transmission distance, $I_{dc}$ is the dark counts per second for the detector, L is the spectral radiance of environment, A is the receiver aperture, $\Delta t^{'}$ is the gate time, h is the Planck constant, c is the speed of light in vacuum, Δλ is the filter spectral width, and Ω = 2π (1 − cos (γ/2)), which is the solid angle of the field of view(FOV), γ is the FOV of the receiver. There are four detectors used in a BB84 system and divided into two groups (each group consists two detectors). Each signal arrives, the data of only one of the four detectors will be treated as the signal. So the wrong bits induced by dark counts and background light are $I_{dc}$ and $\frac{LA\Delta t' \lambda \Delta \lambda \Omega}{4hc\Delta t}$, the total bits (the data of one group for BB84 protocol) of dark counts and background light are $2I_{dc}$ and $\frac{LA\Delta t' \lambda \Delta \lambda \Omega}{2hc\Delta t}$.

In Eq. 1, $\frac{LA\Delta t' \lambda \Delta \lambda \Omega}{4hc\Delta t}$ is the detected photon number of background light for each detector, which ignores the effect of the detector efficiency. The efficiency include the quantum efficiency η of the detector and the transmittance $\eta_{opt}$ of optical elements in the receiver. So the detected background light will be $\frac{LA\Delta t' \lambda \Delta \lambda \Omega \eta \eta_{opt}}{4hc\Delta t}$. Besides, scattering in the ocean and its effects on the polarization of scattered photons should also be considered. Then the sources of wrong bits will include: depolarized photons caused by optical elements, depolarized photons due to scattering, dark counts and detected background light. So, the QBER formula (Eq. 1) should be modified to

$$QBER = \frac{P\frac{\mu\eta\eta_{opt}}{2\Delta t}e^{-\chi_c r} + P_s N + I_{dc} + \frac{LA\Delta t' \lambda \Delta \lambda \Omega \eta \eta_{opt}}{4hc\Delta t}}{P\frac{\mu\eta\eta_{opt}}{2\Delta t}e^{-\chi_c r} + N + 2I_{dc} + \frac{LA\Delta t' \lambda \Delta \lambda \Omega \eta \eta_{opt}}{2hc\Delta t}},$$
(2)

where N is the number of the scattered photons received by the detector, $P_s$ is the probability of the scattered photons that cause errors, and ηopt is the transmission of optical elements. The influence of the scattered photons has been studied in [14] through Monte Carlo method, and the results indicate that the majority of the scattered photons could be well filtered by small FOV and aperture. So the change in the QBER caused by the photon scattering is extremely small (about $10^{-7}$~$10^{-6}$), and is negligible compared to the following results. In the following sections, we will detailed analyze the QBER caused by the background light in different types of water, the optical elements on polarization, respectively.

## 3. ANALYSIS OF BACKGROUND LIGHT

Underwater background light arises mainly from direct incidence and reflection of sunlight, which vary with time, location, turbidity of seawater, detection direction, etc. The intensity of the background light is described by the spectral radiance, which indicates the energy emitted from the unit area of a surface radiation source in unit solid angle and unit time [16, 22]. Previous analyses of the performance of underwater QKD generally used the total irradiance (all the background light in the visible wavelength) to calculate the QBER [12–14]. However, the narrow bandpass filter will eliminate most of the background light. So, only the light with almost the same wavelength as the signal would pass through the filter and induce QBER. Thus, we will calculate the spectral radiance of the underwater environment. Long distance effective key distribution is limited by severe attenuation in turbid seawater. However, there is a blue-green optical window of seawater [16]. Jerlov type seawater is a type of clear open ocean water with subtypes I, II, and III to describe its changes with turbidity. Jerlov type-I is the clearest ocean water, and Jerlov type-II is intermediate [12]. Thus, we calculate the spectral radiance at a wavelength of 480nm in three different radiation directions (upward, downward, and horizontal) at night for several lunar phase angles in Jerlov type seawater. In upward mode, the propagation direction of the quantum signal is upward, the receiver faces downward and is immovable (here, 1 m below the sea surface), and the signal transmitter is moved downward to change the propagation distance. In downward mode, the signal propagates downward. The transmitter remains stationary, and the receiver moves downward. In horizontal mode, the signal transmitter and receiver are located 100 m below the sea surface, and the signal is propagated along the horizontal direction. When the background light is calculated using Hydrolight, the phase function we select is "average particle," and the bottom model is "average seagrass." The average particle phase function is estimated on the basis of the measured data and is adequate for many radiative transfer calculations [16]. As examples of Jerlov type-I and Jerlov type-II water, we calculate the Mid-Pacific Ocean and Northern Pacific Ocean [12], respectively. Thus, the depth of the sea floor is 4000 m, which is the average depth of the Pacific Ocean [23]. The results are illustrated in Fig. 1. Fig. 1(a), (b), and (c) show the spectral radiance of the background light as a function of depth in Jerlov type-I seawater, with an attenuation coefficient of 0.03/m, for different lunar phase angles (full moon, gibbous moon, and quarter). The results for Jerlov type-II seawater, with an attenuation coefficient of 0.18/m, are presented in Fig. 1(d), (e), and (f). The spectral radiation clearly decreases with increasing depth, and the spectral radiance is maximum under full moon conditions.

Among the threemodes, the downward background light has the highest spectral radiance magnitude, which is approximate two or three orders higher than that of the other two modes. The downward light is the downward component of the environmental light incident on the water, and the horizontal and upward light are generated from environmental light by scattering by particulate matter in the water and reflection from the sea floor. For clear ocean water, the scattered light level is low compared to the total light. In general, for the Jerlov type-I seawater, the spectral radiance decrease form approximate $10^{-6}$ $W/(m^2 \cdot sr \cdot nm)$ to $10^{-9}$ $W/(m^2 \cdot sr \cdot nm)$ for downward mode and it decreases from approximate $10^{-9}$ $10^{-8}$ $W/(m^2 \cdot sr \cdot nm)$ to $10^{-13} \sim 10^{-11} W/(m^2 \cdot sr \cdot nm)$ or smaller for the other modes, which will be different for different moon phase angle. For the Jerlov type-II seawater, which has a attenuation coefficient 0.18/m, the spectral radiance will be with a smaller level.

### 4. EFFECT OF OPTICAL COMPONENTS

Besides background light, optical elements will also induce QBER for underwater QKD. However, how a group of optical elements with imperfection affect the QBER for a BB84 protocol underwater QKD system has never been investigated. In this section, as a necessary step to calculate the QBER for underwater QKD, we will study how the imperfect optical elements affect the polarization. In the polarization-encoded BB84 protocol, four polarization states will usually be combined by a beam splitter (BS) at the transmitter site and split at the receiver site. In this paper, the four states we consider are Horizontal(H), Vertical(V), $45°$ (D) and $135°$ (M). Considering the four types of polarizations produced by imperfect polarizer and wave plate, in this section, we analyze the influence of beam splitter, imperfect polarizer and wave plate on polarization, and calculate the QBER.

The relationship between the stokes parameters of the incident light beam $I_0$ and the emerging light beam $I_1$ after passing through an optical components can be described by a 4×4 Mueller matrix [20].

For the BB84 protocol shown in Fig. 2, the H and V states are transmitted through the BS and the D

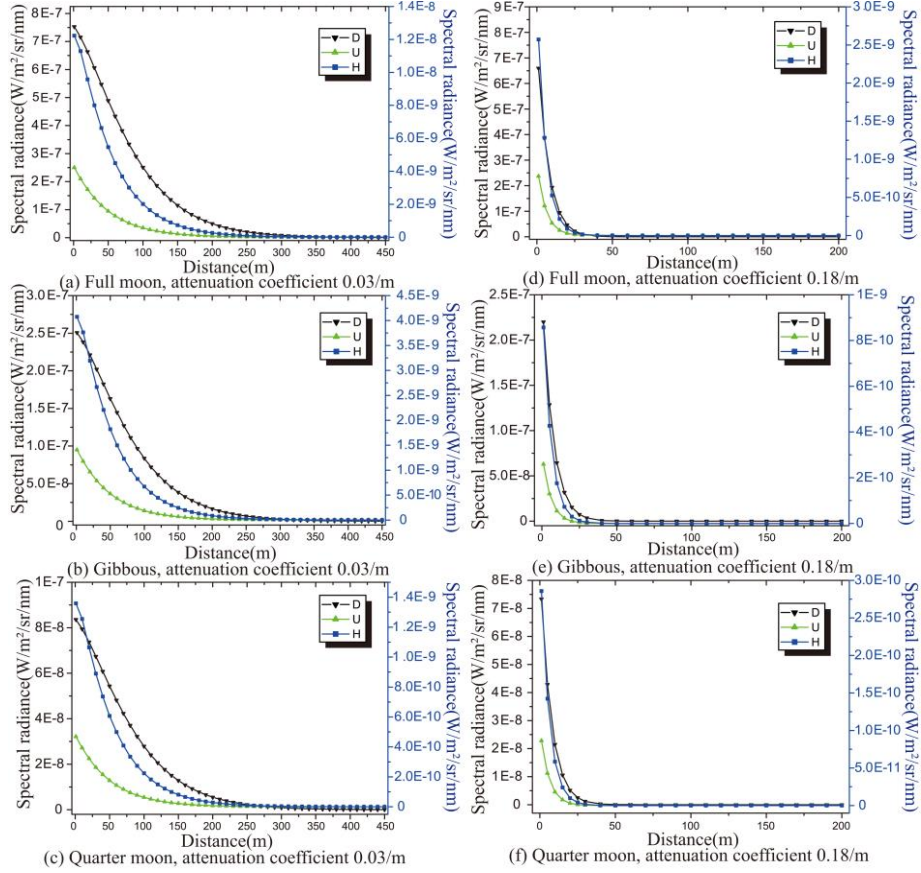

**Fig. 1.** (color online) Spectral radiance curve of underwater background radiance in Jerlov type seawater with respect to different lunar phase angles. The black down triangle (see the left axis), green up triangle, blue box (see the right axis) stand for downward(D), downward(D) and horizontal(H) spectral radiance, respectively.

and M states are reflected, both at the transmitter and receiver. Then the H and V polarized states reaching the detector follow $I_1 = M_{p2} M_t M_t M_{p1} I_0$ where where $M_t$ represents the Mueller matrix of the beam splitter for the transmitted light, $M_{p1}$ and $M_{p2}$ represent the Mueller matrices of the polarizer of the transmitter and receiver with an extinction ratio, respectively. The 4 × 4 Mueller matrix of a polarizer with a extinction ratio of $\varepsilon^2$ is

$$\begin{aligned} M_{11} &= 1+\varepsilon^2 \\ M_{22} &= (1+\varepsilon^2)cos^2\, 2\theta + 2\varepsilon\, sin^2\, 2\theta \\ M_{33} &= (1+\varepsilon^2)sin^2\, 2\theta + 2\varepsilon\, cos^2\, 2\theta \\ M_{44} &= 2\varepsilon \\ M_{12} &= M_{21} = (1-\varepsilon^2)cos\, 2\theta \\ M_{13} &= M_{31} = (1-\varepsilon^2)sin\, 2\theta \end{aligned} \quad (3)$$

whose other elements are 0. The D, M polarization states reaching the detector follows $I_2 = M_{p2} M_{hp2} M_r M_r M_{hp1} M_{p1} I_0$, where $M_{hp1}$ nd $M_{hp2}$ represent the Mueller matrices of a half wave plate of the transmitter and receiver with a retardation accuracy, $M_r$ represents the Mueller matrix of the beam splitter for the reflected light

The Mueller matrix of a BS for the transmitted light $M_t$ is

$$\begin{pmatrix} \frac{1}{2}(t_p + t_s) & \frac{1}{2}(t_p - t_s) & 0 & 0 \\ \frac{1}{2}(t_p - t_s) & \frac{1}{2}(t_p + t_s) & 0 & 0 \\ 0 & 0 & \sqrt{t_p \cdot t_s}\, cos\phi_t & -\sqrt{t_p \cdot t_s}\, sin\phi_t \\ 0 & 0 & \sqrt{t_p \cdot t_s}\, sin\phi_t & \sqrt{t_p \cdot t_s}\, cos\phi_t \end{pmatrix} \%,$$

(4)

and the Mueller matrix of a BS for the reflected light $M_r$ is

$$\begin{pmatrix} \frac{1}{2}(r_p+r_s) & \frac{1}{2}(r_p-r_s) & 0 & 0 \\ \frac{1}{2}(r_p-r_s) & \frac{1}{2}(r_p+r_s) & 0 & 0 \\ 0 & 0 & \sqrt{r_p \cdot r_s}\cos\phi_r & -\sqrt{r_p \cdot r_s}\sin\phi_r \\ 0 & 0 & \sqrt{r_p \cdot r_s}\sin\phi_r & \sqrt{r_p \cdot r_s}\cos\phi_r \end{pmatrix} \%.$$

(5)

with transmissivity(reflectivity) of where $tp(rp)$ and $rp(rs)$ represent the transmissivity (reflectivity) for the p and s polarization states, and $\phi t(\phi r)$ represents the phase difference of the p and s states for transmitted (reflected) light. The Mueller matrix of a wave plate is

$$M_{11}=1, M_{22}=\cos^2 2\theta + \cos\delta\sin^2 2\theta$$
$$M_{33}=\cos^2 2\theta \cos(\delta) + \sin^2 2\theta, M_{44}=\cos\delta$$
$$M_{23}=M_{32}=\cos 2\theta \sin 2\theta - \cos 2\theta \cos\delta \sin 2\theta.$$
$$M_{24}=-M_{42}=-\sin 2\theta \sin\delta$$
$$M_{34}=-M_{43}=\cos 2\theta \sin\delta$$

(6)

where δ is the phase difference between the fast and slow axis, and θ is the angle of the fast axis, other elements are 0. Typically, for a half wave plate, δ = π. In practical applications, the extinction ratio of a polarizer can reach 1:10000 [21], and the retardation accuracy of a wave plate can be limited within λ/300 [22, 23]. The rotation accuracy of the polarizer and the wave plate is 5 arc minutes, which is easy to realize for the optical mounts (such as PRM1 of Thorlabs).

Generally, the H, V polarization states are minimally influenced by a BS. Because the influence of a BS on the polarization states is due to the different transmission(reflection) and the phase difference of the p and s states. In the ordinary case, the $r_p$ and $r_s$ are both 50 ± 5% and the $\phi_r$ s about $7°-9°$, then P ≈ 0.017. For the optimized non-polarizing beam splitters, the $r_p$ and $r_s$ can reach 50 ± 0.5%, and $\phi_r = (0\pm0.3)°$ [24], resulting to P ≈ 2.3 × 10−4. Also, we will analyze the QBER and key rate under these two situations, an ordinary case and an optimal case, in the following section

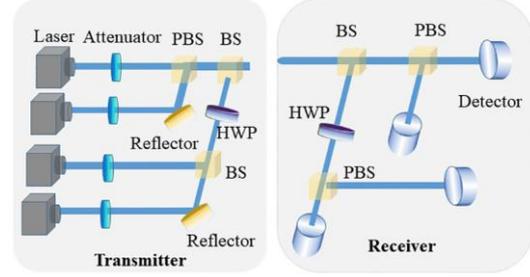

**Fig. 2**. . (color online) Sketch of the transmitter and receiver of a BB84 protocol, the polarized photons will output from the transmitter, propagate through the channel and input into the receiver. HWP represents the half wave plate, PBS represents the polarization Beam Splitter.

## 5. PERFORMANCE ANALYSIS OF UNDERWATER QKD SYSTEMS

The QBER is an important performance indicator of QKD. In this section, we first analyze the main factors affecting the QBER of underwater QKD systems. Then, the QBER for various distances is studied for the upward, downward, and horizontal transmission modes. Finally, the key rates in the three modes are calculated, including the sifted key rate and final key rate.

### A. Analysis of the QBER of underwater QKD

According to Eq. 2, there are three main factors, i.e., the dark counts, the background light and the misalignment of optical components, contributing to the QBER. The dark counts (Idc) and the detection efficiency ($\eta$) of single photon detectors directly determine the quantum key generation rate and distribution distance. Idc, $\eta$ are 100 counts/s, 20% for an ordinary single photon detector and 1 count/s, 80% for the superconducting nanowire single-photon detector. The background light is determined by the environment, as illustrated in Fig. 1.

The system parameters used for calculating the QBER and key rate of the ordinary and optimal cases are listed in Tables 1 and 2, respectively. The same parameters are not listed in Table 2. P is calculated in section 4, $\chi_c$ is the attenuation coefficient of the Jerlov type water, Δλ in Table 1 is the bandwidth of an ordinary filter(such as the filter made in Thorlabs) and Δλ in Table 2 is from Ref. [13], f is the frequency of the quantum signal [25], $\eta_{opt}$ is evaluated according to the transmissivity of the optical elements of the receiver in Fig.6, A is from Ref. [13].

**Table 1. System parameters for the ordinary case**

| $P = 0.017$ | $\Delta\lambda = 1nm$ | $\eta = 20\%$ |
|---|---|---|
| $\Delta t' = 5ns$ | $I_{dc} = 100$ | $f = 40MHZ$ |
| $\gamma = 10mrad$ | $\mu = 0.1$ | $\chi_c = 0.03, 0.18/m$ |
| $\eta_{opt} = 95\%$ | $A = 30cm^2$ | |

**Table 2. System parameters for the optimal case**

| $P = 2.3 \times 10^{-4}$ | $\Delta\lambda = 0.12nm$ | $\eta = 80\%$ |
|---|---|---|
| $\Delta t' = 200ps$ | $I_{dc} = 1$ | |

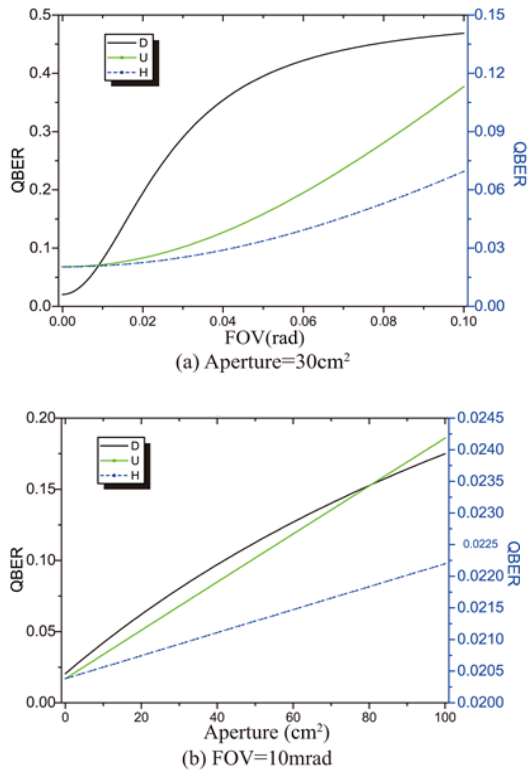

**Fig. 3.** (color online) (a). QBER as a function of FOV (with a fixed aperture of 30cm2 ). (b). QBER as a function of aperture (with a fixed FOV of 10 mrad). The black line (left axis) describes the downward mode and the blue and green lines (right axis) describe the other two modes. The transmission distance of the signal is 100 meters. Other system parameters are shown in Table 1.

We first analyze the QBER caused by the background light. For a given condition, the background light received by the detector depends mainly on the propagation direction, receiver aperture and FOV. The effects of the receiver aperture and FOV under full moon conditions in Jerlov type-I seawater are shown in Fig. 3 (a) and (b), respectively. Obviously, the QBER increases as the FOV and receiver aperture increase. Fig. 1 (a) shows that the intensity of the background light is maximum for the downward mode. Secure QKD can be achieved when the FOV is less than 11mrad. For the upward and horizontal modes, the QBER increases slowly when the FOV is in the range of 0 to more than 10mrad. When the FOV reaches dozens of milliradians, the QBER gradually becomes more sensitive to the increase in FOV. The background light becomes the main factor affecting the QBER. As shown in Fig.3 (b), the correlation between the QBER and receiver aperture is approximately linear, which is especially obvious for the upward and horizontal modes. Thus, we can effectively reduce the QBER induced by the background noise by choosing a smaller FOV, and a small aperture is also helpful for reducing the QBER.

To decrease the influence of the background light on underwater QKD, a small FOV (10mrad) and aperture (30cm²) in the secure QKD range are selected to investigate the relationship between the QBER and the propagation distance. Fig 4 shows the QBER in Jerlov type-I seawater caused by the optical elements optical elements( $Q_{opt}$ ), the background light( $Q_{bac}$ ) and the dark counts( $Q_{I_{dc}}$ ) under full moon condition. moon condition. The

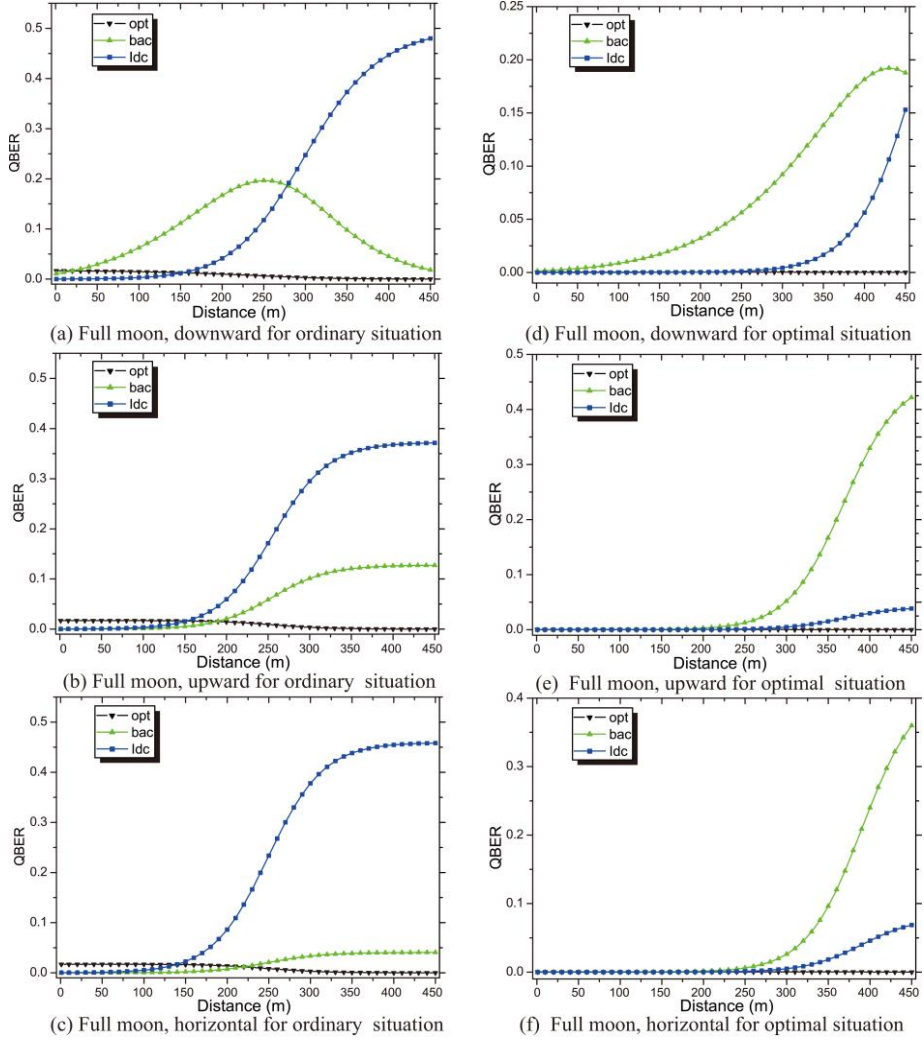

Fig. 4. (color online) QBER come from different factors, Qopt, Qbac, QIdc as a function of propagation distance under full moon condition with an attenuation coefficient of 0.03/m. (a), (b), (c) are calculated according the parameters shown in Table 1 and (d), (e), (f) are calculated according the parameters shown in Table 2

QBERs in the ordinary case (Table.1) are shown in Fig.4 (a), (b), (c). The calculation results show that the imperfect optical elements will not have significant impact on underwater QKD. For the underwater QKD within a distance of 150m for upward and horizontal modes, the Qopt, QIdc and Qbac are with almost the same level. As the distance improves, the dark counts will be the main source for the ordinary system according to the results because the Idc will keep constant and signal will attenuate as increasing distance. Note that Qbac in Fig.4 (a) gradually increases and then decreases with increasing distance and the effect of the dark counts will be more obvious at distances greater than 250 m, because attenuation of the background light is slightly lower than that of the signal. The QBERs in the optimal case (Table.2) are illustrated in Fig.4 (d), (e), and (f). When the distance is less than 300 m, which is within the maximum estimated distance for underwater quantum communication [15, 30], $Q_{opt}$ and $Q_{I_{dc}}$ can almost be ignored in the three propagation modes, and the background light will be the main factor of the QBER. Generally, the main sources of the QBER in long-distance QKD are the background light and dark counts. The QBER of the optimal system is much smaller than that of the ordinary system because both dark counts and detected background light are smaller than those of the ordinary system. The low dark counts and background light make it possible for long distance (i.e 300m) QKD in the clearest ocean water for optimal system. For an underwater environment, reducing the dark counts and background light can reduce the QBER effectively, which is useful for remote QKD.To determine the distance of secure QKD, we further calculate the total QBER for the two types of water with differenttransmission modes under full condition. The results are shown in Fig. 5 and 6. When the attenuation coefficient is 0.03/m, the QBER will reach the up boundary of QBER for secure QKD (11% [31]) when the distance is approximate 130 m for the downward mode and approximate 200 m for the other two modes using the parameters in Tab. 1. If we use Eq.1 to calculate the QBER, the QBER will reach 11% when the distance is about 35m for the downward mode and about 190m for the other two modes. That's because the background light of the downward mode is the maximum of the three modes and and Eq.1 will over estimate the QBER caused by background light. Thus, using the modified formula to calculate the QBER is necessary, especially when the background light is severe. The distance can exceed 300m if we improved the performance of the system (Tab.2). See Figs. 5 (a) and 6 (a) for more details. In contrast, for the seawater with an attenuation coefficient of 0.18/m, the secure QKD distance is only a few tens of meters in both cases because of severe attenuation. The downward mode obviously has the highest QBER among the three modes, because the background light is strongest when secure underwater QKD is performed.

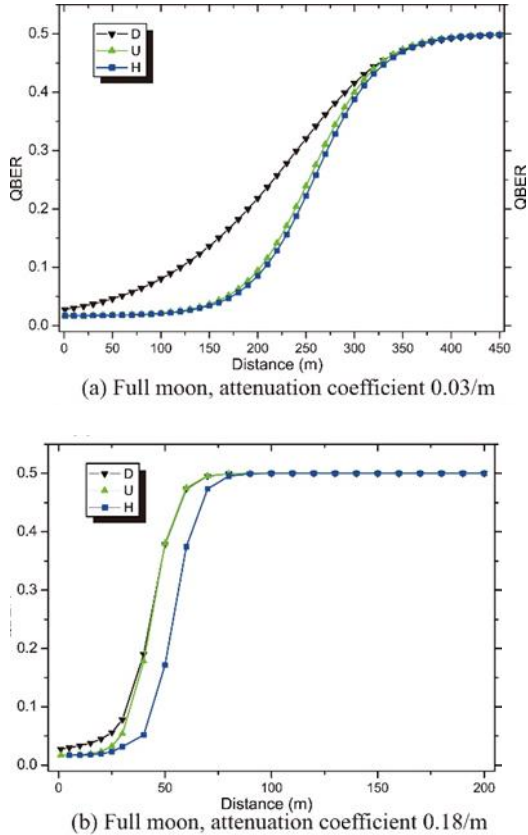

Fig. 5. The total QBER of underwater QKD in Jerlov type seawater for the ordinary case with the different attenuation coefficient. (a). $\chi_c$ = 0.03/m. (b). $\chi_c$ = 0.18/m. The parameters we used to do the calculation are shown in Table 1.

**B. Analysis of the key rate of underwater QKD**

In BB84, through public discussion Alice and Bob reject the key bits where they used different bases; the remaining key, for which their bases agree, is called the "sifted key." Because the secure final key is extracted from the sifted key, which will be determined by the parameters of the system, we calculated the sifted key for underwater QKD for the ordinary system (Tab. 1) and optimal system (Tab. 2) as the preceding step towards calculating the secure key rate. In BB84 protocol, the sifted key rate is [20]

$$k = f \cdot \mu \cdot T_{link} \cdot q \cdot \frac{\eta}{2} \cdot \eta_{opt}, \quad (7)$$

where f is the pulse frequency of the laser, $T_{link}$ is the transmission ratio of seawater channel which obeys Beer-Lambert law [25, 32], $q$ is the sifting factor which is usually $\leqslant 1$ and typically 1 or 1 2, and k is the rate of sifted key. In this paper, the value of q is taken as 1. With the QBER shown in Fig.5 and 6, we can obtain the sifted key rate of underwater QKD, which is illustrated in Fig. 7.

The sifted key rate for the three modes are the same because the parameters in Eq. 7 are independent of the direction of transmission. As shown in Fig 7, the sifted key rate of underwater QKD is approximate 18.9 kbits/s at a distance of 100m in the ordinary case and can reach around 76 kbits/s in the optimal case.

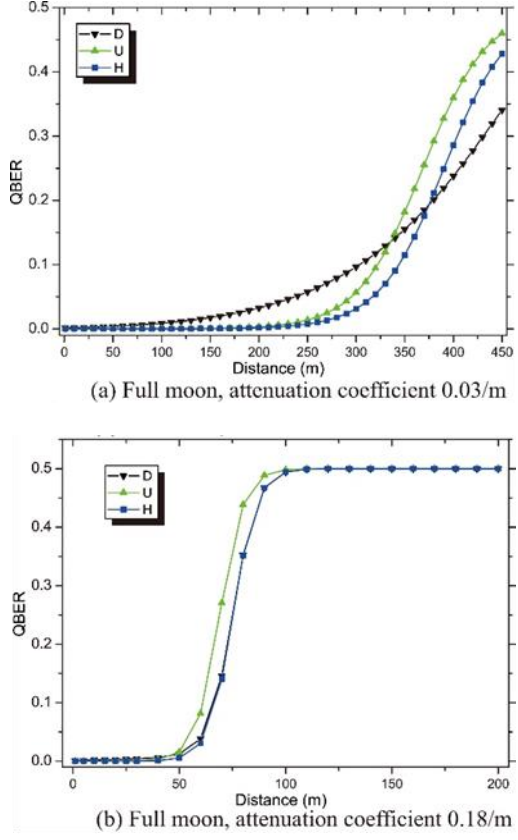

Fig. 6. (color online)The total QBER of underwater QKD in Jerlov type seawater for the optimal case with the different attenuation coefficient. (a). $\chi c = 0.03/m$. (b). $\chi c = 0.18/m$. The parameters we used to do the calculation are shown in Table 2.

Then, the final key rate for practical QKD with decoy state can be estimated according to the GLLP formula [29, 30]:

$$R \geqslant q[-Q_\mu f(E_\mu)H_2(E_\mu) + Q_1[1-H_2(e_1)]], \quad (8)$$

where $Q_\mu$ is the gain of signal states, $E_\mu$ is the QBER of singal states, $Q_1$ is the gain of single-photon states, $e_1$ is the error rate of single photons, $f(E_\mu)$ is the error correction efficiency and $f(E_\mu) \geqslant 1$, $H_2(x) = -x\log_2(x) - (1-x)\log_2(1-x)$. is the binary Shannon entropy. For the QKD without decoy states, the final key will be [29, 30]:

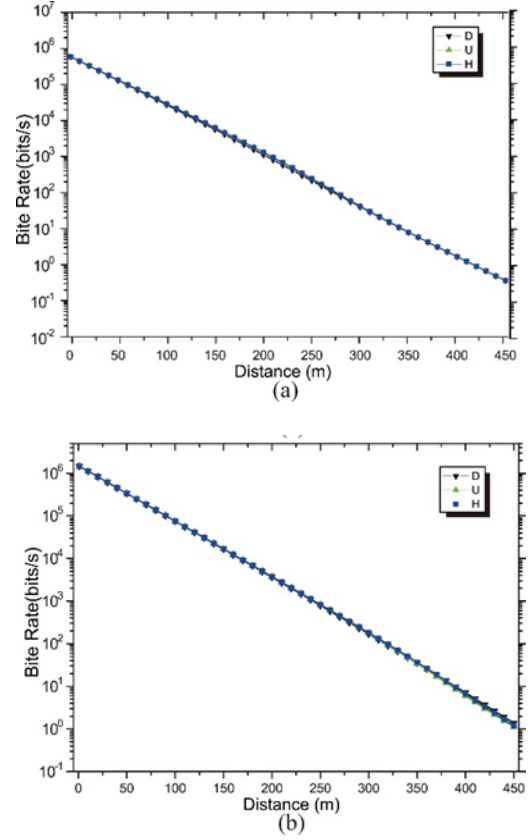

Fig. 7. The sifted key rates as a function of propagation distance in clear Jerlov type-I seawater, under full moon condition. The lines with up triangles, down triangles and squares represent upward, downward and horizontal communication modes, respectively. (a) is calculated according the parameters shown in Table 1 and (b) is calculated according the parameters shown in Table 2.

$$R \geqslant Q_\mu \cdot [-f(E_\mu)H_2(E_\mu) + \Omega(1-H_2(E_\mu/\Omega))], \quad (9)$$

where $\Omega$ is the fraction of "untagged" photons. Then the final key rate in clearest Jerlov type I seawater can be calculated, which is shown in Fig. 8.

For the ordinary system, the security key rate is approximate 1.8kbits/s when the propagation distance is 100 m for the downward mode and approximate 8kbits/s for the other two modes. For the optimal optical parameters, the key rate can reach 57.2kbits/s for downward mode and about 67kbits/s for the upward and horizontal modes at a distance of 100m. The maximum secure distance can reach about 310m, 320m and 340m for downward, upward and horizontal modes, respectively. For practical

QKD with decoy state, the final key rate, which can be estimated according to Eq. 8, will be improved.

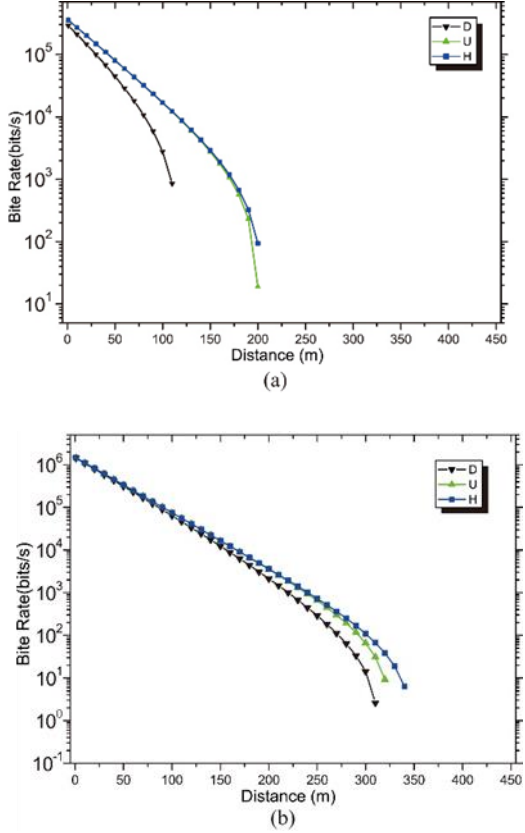

Fig. 8. Final key rate of underwater QKD in the ordinary case (a) and in the optimal case (b).(a) is calculated according the parameters shown in Table 1 and (b) is calculated according the parameters shown in Table 2

Considering the simple onedecoy protocol [30], we have $Q_1 = \mu e^{-\mu} Y_1, E_1 = \dfrac{E_\nu Q_\nu e^\nu}{Y_1 \nu}$, where $Y_1 = \dfrac{\mu}{\mu\nu - \nu^2}(Q_\nu e^\nu - Q_\mu e^\mu \dfrac{\nu^2}{\mu^2})$, μ and ν represent the mean photon number per pulse of signal state and decoy state, respectively. (a) Full moon, attenuation coefficient 0.03/m (b) Full moon, attenuation coefficient 0.18/m Fig. 6. (color online)The total QBER of underwater QKD in Jerlov type seawater for the optimal case with the different attenuation coefficient. (a). $\chi_c$ = 0.03/m. (b). $\chi_c$ = 0.18/m. The parameters we used to do the calculation are shown in Table 2. We calculate the final key for the ordinary case according to Table.1 under full moon condition when μ = 0.48 and ν = 0.05. The results show that the final key rate can reach 20.3, 32.8 and 32.7kbits/s at a distance of 100m for the upward, downward and horizontal modes, respectively.

## 6. CONCLUSION AND DISCUSSION

In this paper, we first modified the formula used to calculate the QBER. Then we analyzed the influence of imperfect optical elements on the polarization states and studied the background light underwater in detail using Hydrolight. Finally, we investigated the performance of underwater QKD. Using the modified formula and the background light calculated by Hydrolight, we found that the sifted bit rate and secure bit rate for QKD in Jerlov type-I water are 18.9 and 8kbits/s under full moon condition at a distance of 100m, respectively, and the maximum distance can reach 200m for the ordinary system. For the optimal system, the sifted bit rate and secure bit rate for QKD in Jerlov type-I water are 76 kbits/s and 67kbits/s under full moon condition at a distance of 100m, respectively, and the maximum distance can exceed 300m. Previous analysis[14] on underwater QKD, which used Eq.1 and the total irradiance of all visible light to estimate the QBER, showed that QKD in Jerlov type-I seawater can be performed at a distance of more than 100m under star only environment, and QKD cannot be performed under full moon condition because the background light is so severe that the QBER will exceed 25%. However, using Eq.1 and the total irradiance to estimate the QBER will overestimate the QBER induced by background light.

Although the final key rate is very low in this study, it can be increased by increasing the pulse repetition rate f . Typically, for a semiconductor laser with a wavelength in the blue-green window, the frequency of it can reach 100-200MHZ. The count rate of a superconducting nanowire single-photon detector can also reach more than 100MHZ. The performance of underwater QKD can also be improved by applying decoy states, and it is also important to improve the performance of the polarization elements and single photon detector. With increasing the final key rate and secure communication distance, practical application of underwater quantum communication will become possible in the future. The attenuation coefficient is affected by wavelength and the light we analyzed in this paper has a wavelength of 480 nm, which is in the blue-green optical window of seawater. If another wavelength is selected, the results will be different and can be calculated accordingly. Further, the type of seawater will also affect the attenuation coefficient because the main component will differ for different types of seawater and the absorption peak willvary depending on the component. The results show that improving the performance of the optical elements, especially the detector, will be helpful for implementing underwater QKD. Further, the synchronous signal, which is used to control the detect to work once the signal is arriving, is also necessary for underwater QKD. If photons arriving in the same detection window, the data will be abandoned or used to check wherther blinding attacks exists in the QKD. To avoid the effect of the synchronous signal on the QBER, another wavelength, also in blue-green range, is necessary and the time delay between quantum signal and synchronous signal is also essential. In practical underwater environments, many other factors will affect the propagation of the beam light for underwater QKD, such as dot whipping caused by turbulence, and wavefront distortion. These factors may cause depolarization and extra loss, which will effect the QBER and key rat accordingly. However, the results here are still available when a stable communication link between Alice and Bob is established and the reference system between them are coincident. To fulfill the requirements, the problems affecting the propagation of beam light need to be solved by referring to the techniques in free space QKD and optical communication, such as ATP [10, 11] and adaptive optics system [35].

## ACKNOWLEDGMENTS

This work was supported by the National Natural Science Foundation of China (Grant Nos. 61575180, 61701464 and 11475160 ).